# LAC : LSTM AUTOENCODER with Community for Insider Threat Detection

Sudipta Paul

National Institute of Science Education and Research Bhubaneswar
Khurda, Orissa, India
Homi Bhaba National Institute
Anushaktinagar, Mumbai, India - 400094
sudiptapaulvixx@niser.ac.in

Subhankar Mishra

National Institute of Science Education and Research Bhubaneswar
Khurda, Orissa, India
Homi Bhaba National Institute
Anushaktinagar, Mumbai, India - 400094
smishra@niser.ac.in

**ABSTRACT**

The employees of any organization, institute or industry, spend a significant amount of time on computer network, where they develop their own routine of activities in the form of network transactions over a time period. Insider threat detection involves identifying deviations in the routines or anomalies which may cause harm to the organization in the form of data leaks and secrets sharing. If not automated, this process involves feature engineering for modeling human behavior which is a tedious and time-consuming task. Anomalies in the human behavior are forwarded to a human analyst for final threat classification. We developed an unsupervised deep neural network model using LSTM AUTOENCODER which learns to mimic the behavior of individual employees from their day-wise time stamped sequence of activities. It predicts the threat scenario via significant loss from anomalous routine. Employees in a community tend to align their routine with each other rather than the employees outside their communities, this motivates us to explore a variation of the AUTOENCODER, LSTM AUTOENCODER- trained on the interleaved sequences of activities in the Community (LAC). We evaluate the model on the CERT v6.2 dataset and perform analysis on the loss for normal and anomalous routine across 4000 employees. The aim of our paper is to detect the anomalous employees as well as to explore how the surrounding employees are affecting that employees' routine over time.

## 1 INTRODUCTION

A very well known problem in security analytics is insider threat detection [11, 16, 18]. Insiders are persons within the organizations trusted with sensitive and personal information. Insider threat detection is the process of finding out potential threats through their unauthorized actions leading to damage to the organizations such as data leaks. The general approach is to view it as anomaly detection problem through the user logs which qualifies all of the properties especially volume and variety of big data with real-time streaming.

**Anomaly** can be defined as an outlier in a set of dataset which affects the entirety in an unacceptable way. In our case it can be defined as the fluctuation from the normal actions and activities over a time period. According to Hawkins[9], the definition of an outlier would be " an observation which deviates so much from other observations as to arouse suspicions that it was generated by a different mechanism". According to Suri et al.[15]," An object in a data set is usually called an outlier if- (1) It deviates from the normal/known behavior of the data, (2) It assumes values that are far away from the expected/average values, or (3) It is not connected/similar to any other object in terms of its characteristics. Therefore, (1) it can make an individual assume a leadership role and influence others to follow e.g. lateral thinking is one such case, (2) It may play a disturbing role in summarizing the overall behavior of a community of people/entities." It is entirely data, place and situation centric. So, the time frame is an important player to detect an outlier behavior.

These behaviors can be good or bad depending on the sequences and aftermath of situations. We consider only the situations where bad consequences happen as a result, as an anomaly. Some possible outlier and anomalous situation in an organization might be of concerns are - unsanctioned data has been transferred, sabotage of resources has been occurred, When the employee stops working, How often any employee mails to the people who are outside of the domain of the company, How often any employee uses their other mail-IDs(not the one provided from the company domain), What is the frequency of using VPN, How frequently the sent mail has an attachment, How many times the employees login to the their machines, How many times the employees visit to certain websites, How many times the employees access to the shared file in the network to download, upload or for other activities, Most importantly how these aforementioned situations are related. An outlier or anomalous situations can result to a better profit for an organization whereas the opposite is an option too. It is almost impossible for a human to keep a track of all these situations at once, as all these data are in large volumes with high velocity and veracity.

**Importance of Trust**. When anomalies might be a potential threat to the organizations, and comes from the employees themselves, detection of such situations is known as insider threats detection. Detection of anomalies is extremely complex and challenging. For example - in a hypothetical situation one of the potential insider threat employees can perform an unauthorized task by the use of his trusted access in the Intranet of the institution. Therefore the external network security devices won't be able to detect them. In another situation the harmful employee might hold a grudge against another employee that encourages him to plant a logic bomb to the second employee's machine to steal intellectual property or destroy the whole system [18]. The basis of all these situations is the trust that the employee has achieved through his course of time and work routine in the institution. Using this trust, to hide his/her detection the employee can make sure that the abnormal activities over time, will be distributed in the log line records to increase the difficulty for the analyst! Hence, modeling of behavior of individuals would alone not suffice for Insider Threat Detection [16].

**Importance of Data**. All the above situations should be understood at once to classify them as anomalies. According to Aggarwal[1], "Virtually all outlier detection algorithms create a model of the normal patterns in the data, and then compute an outlier score of a given data point on the basis of the deviations from these patterns."For example, nearest neighbor-based outlier detection algorithms model the outlier tendency of a data point in terms of the distribution of its k-nearest neighbor distance. Thus, in this case, the assumption is that outliers are located at large distances from most of the data. But, most of these models are based on supervised learning, and score. [16]

But, the problem itself has an inherent unsupervised nature. This is the baseline of the unsupervised deep learning models for anomaly detection. They are well known for their ability to extract hidden patterns in a large amount of data. Here the extracted log lines from the network transactions dataset contain sequences of day, timestamps and activity with different categorical and attribute features in interleaved manner. We will explore this topic further in the related work section 2.

**Importance of Community.** The theoretical study of the socialize phenomenon of the human race is known as "Group Dynamics", a term coined by Kurt Lewin[5, 17] in 1943. The "Thomas theorem"[5, 17], applied to group suggests that if individual thinks an aggregate is a true group, then the group will have interpersonal consequences for those in the group and for those who are observing it. This is the power of a group or community. An organization, institute or company is a sum of different communities on the basis of different attributes that abide the social conformity. These communities have members not only from a single organization but from different organizations. The employees other than the concerned organization in a community are denoted as "outsiders". The members of a community have the ability to do harm towards the organization by working together or individually with the direct or indirect help of the members. According to Forsyth[5], " A group or Community is two or more individuals, who are connected by and within social relationship." A community or group seeks various goals: generating, choosing, negotiating and executing which in turn create inter dependence among the group members. Early works of Tripllet and Milgram[5] shows that how one person of a group or community can influence others and how detrimental the effect can be. Here we extracted the communities on the basis of Louvain algorithm using the email log lines. Details of the whole process will be discussed in the subsection 4.3. Using the proposed model, our aim is to study, how insider threat happens by a single employee and what the influence of the community on that employee is. This is discussed in the subsection 4.4.

**Importance of deep learning models.** Deep learning networks like CNN, RNN, DBN, DNN etc. are the subset of machine learning algorithm family where they use layered structure to extract higher level features progressively on the given input. Another point to be considered here is that the deep learning models have the inherent ability to understand the temporal behavior inside the log lines, whether it is applied to the individual employee or the sequences of time lines. Deep Neural Network is the straight answer for these situations [13]. This will be explored in detail in the section 2.

Our contributions are -

- Introduction of community detection algorithm to detect the underneath communities with the help of friendship graph.

- Introduction of LSTM AUTOENCODER for the first time to model the insider threat detection.
- Generation of feature set with the granularity of event occurrence with respect to the activities
- Introduction of a regeneration based overall system instead of a score specific one.

In this paper we are going to explore the current state-of-the-art models in section 2. We will explain our solution in details in section 3 and 4. We will share our future plan with conclusive remarks in section 5.

## 2 RELATED WORK AND BACKGROUND THEORY

**Deep Neural Network.** In the paper [13] they have explicitly shown that even though there exists some drawbacks but machine learning approach is the most promising solution to anomaly detection in net-work intrusion problems. They have pointed out the possible cons in terms of high cost of errors, semantic gaps etc. Mitigation measures include knowing the possible threat models, keeping the scopes narrow, reducing the costs by limiting false positive and last and most importantly having an under-standing of data to apply the machine learning algorithms effectively. [4] have given a very good summary of what should be the problem definition and formulation for anomaly detection in sequence data using machine learning and deep learning model in the continuity of the above paper. They recognized three unique anomaly detection problem formulations. They have also categorized the existing techniques with respect to underlying algorithm which are basically variants of the basic algorithms. They described how to produce anomaly score in different anomaly detection situations.

[16] have used two approaches towards this problem - one is deep learning model using vanilla AUTOENCODER and another is stacked Recurrent Neural Network using Long Short Term Memory(lstm) layer. They presented three sets of experiment communities, with the three queries in mind. They are - what is the effect of including or excluding the categorical value in the model input and output, what is the best prediction mode - same time step or in the next time step and what is the effect of co-variance type for the continuous features with a contrast to the baseline models. They modeled the normal behavior in both the approaches and used anomaly as an potential indicator of threats. The cons of their model was they did not take a notice of sequence of activities but the aggregated count of activities, which is way too much generalization for the anomaly detection problem. Also, they didn't explore the granularity of time with respect to the occurrence of activities. We took acknowledge of these shortcomings and hence provided a detailed solution in section 3.

**LSTM.** The full form is: Long short-term memory (LSTM). Learning to store knowledge for an extended time periods by recurrent back propagation is a very lengthy process, most of the time because of inadequate, decaying error back flow. LSTM cell maps an input sequence to a hidden state sequence.

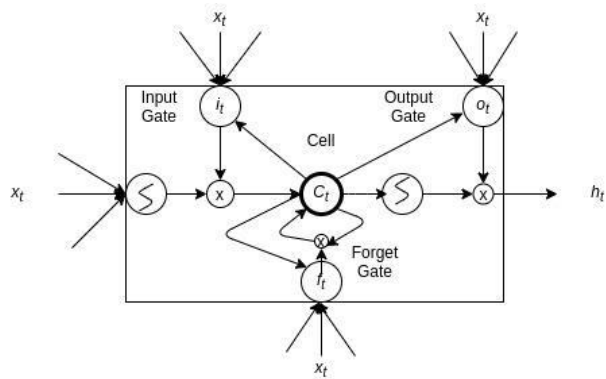

**Figure 1. LSTM Cell Architecture reproduced from [14]**

Here the hidden state is computed as a function of the input sequences and not on the last input alone. Conditioning the hidden layer on a sequence rather than the current input alone allows us to capture temporal patterns in employee behavior further to build an increasingly accurate model of the employee's behavior over time. Long short-term memory (LSTM) [10] architecture is in Figure 1. Here hidden state at a time-stamp is a function of a long term memory cell. In a deep LSTM with hidden layers, our final hidden state, the output of hidden layer depends on the input sequences and cell states that have been given in the [8].

**AUTOENCODER.** An AUTOENCODER [7] is an unsupervised neural network by architecture that is trained to mimic the input pattern in the reconstructed output. Suppose, it consists of a hidden layer $h$ that has a pattern representation of the *input(x)*. The decoder then produces a reconstruction $r = g(h)$.

By Design, AUTOENCODERs are not able to mimic the input patterns as it is. They are restricted in a certain manner that forces them to copy approximately, that somehow takes after the input training data. The encoder part is always forced to align to learn the useful pattern of the input data but not to copy the whole thing. Modern AUTOENCODERs [2] have generalized the idea of the whole architecture beyond deterministic functions to stochastic mappings i.e. $P_{encoder}(h/x)$ and $P_{decoder}(h/x)$.

**LSTM AUTOENCODER.** In this paper [14], they used the LSTM AUTOENCODER frame-work for the first time to learn the representation of video by frame to frame. The Encoder is built from a LSTM layer, is feed by the sequences of video frames to produce an encoded presentation. This encoded presentation is then decoded through a decoder layer made from LSTM to produce again a sequence of frames that might be similar to the input. They considered different options for the output frames. One option was the prediction of similar input frame sequences. The motivation was same as vanilla AUTOENCODERs – to catch all the useful features to reconstruct the initial frames. Another option was the prediction of the succeeding frames. They later combined these to options. The inputs to the model are a single video frames. However, for the evaluation process, they considered only two kinds of inputs. They are image patches and high-level "percepts" extracted by applying a convolutional net pretrained on ImageNet. To test their first hypothesis they used normal image frames with the combination of moving MNIST digits dataset. For the second hypothesis they used percepts which are the states of the last (and/or second-to-last) layer of RELU (hidden) units. The reproduced model from their work is in figure 2. We are using LSTM AUTOENCODER for the first time in anomaly detection analysis.

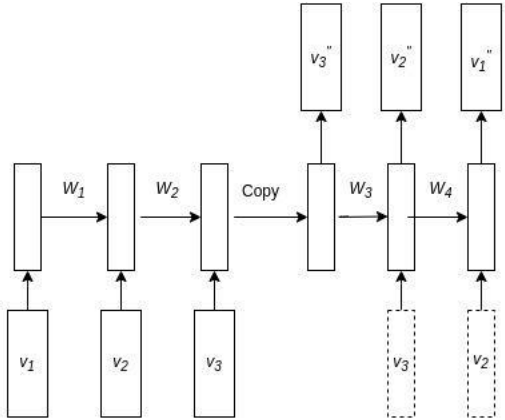

**Figure 2. LSTM AUTOENCODER reproduced from [14]**

**Louvain Algorithm.** According to [3] social, technological and information systems can often be described in terms of complex networks that have a topology of interconnected nodes combining organizations and randomness. On the basis of this interconnection and modularity we can easily deduct different communities in a dataset. We use one of the most popular algorithms, Louvain algorithm [12] here to detect the communities. It is a greedy algorithm that returns communities without overlapping. It works on modularity. Modularity is defined as follows –

$$Q = \frac{1}{2m} \sum_{i,j} \left[ A_{i,j} - \frac{k_i k_j}{2m} \right] \delta(C_i C_j)$$

Here, is $A_{i,j}$ the edge weight between nodes $i$ and $j$ ; $k_i$ and $k_j$ are the sum of the weights of the edges attached to nodes $i$ and $j$, respectively; $m$ is the sum of all of the edge weights in the graph; $C_i$ and $C_j$ are the communities of the nodes; and δ is Kronecker delta function (δ $_{x,y}$ = 1 if $x$=$y$ , 0 otherwise). The algorithm will optimize modularity and according to that modularity it will assign the nodes to their respective communities.

## 3 MODEL DESCRIPTION

Our **LAC (LSTM AUTOENCODER with Community)** model works in the following two consecutive phases.

**First Phase:: Community Detection.** Here, we use the Louvain algorithm. The basis of using this algorithm is that an employee will send a mail only to another person when he knows him and this is indicated by an edge in a "friendship graph"[19]. In their paper [19] Grotto et al. mathematically shows that how a group or community is powerful enough to affect the group dynamics. We extend this notion to hypothesize that people in communities behave similarly and it is easier to spot those involved in anomalous activities in a community. With the help of this friendship graph we then extracted the communities using Louvain algorithm. We introduce this phase to - extract the communities on the basis of friendship graph and modularity, this will also help in introduction of parallel analysis with respect to the communities which in turn helps to reduce the time of analysis in whole and increase the number of employees for online analysis at a time, Visualize and analyze the influences of other employees on the anomalous employees in the later stage. The description of the communities is given later in section 4.

**Second Phase:: LSTM AUTOENCODER RNN model**. Our deep learning model consists of an encoder (Figure 3) and an decoder made of LSTM. Together with the encoder and decoder the LSTM AUTOENCODER works.
We use the encoded tensor from the encoder further to recreate the employees'

```
Model
Model: "model"
_________________________________________________________________
Layer (type)                 Output Shape              Param #
=================================================================
input_1 (InputLayer)         [(None, 6502, 22)]        0
_________________________________________________________________
lstm_1612 (LSTM)             (None, 6502, 128)         77312
_________________________________________________________________
dense_3224 (Dense)           (None, 6502, 64)          8256
_________________________________________________________________
dense_3225 (Dense)           (None, 6502, 22)          1430
=================================================================
Total params: 86,998
Trainable params: 86,998
Non-trainable params: 0
_________________________________________________________________
```

**Figure 3. Description of Encoder.**

normal activities using the decoder part for each and every communities parallely. Therefore whenever an abnormal sequence of activities (which are not following the normal behavior of individual with respect to its community) will occur the reconstruction loss will be higher. The description of the encoder model is in figure 3.

The description of the whole LSTM AUTOENCODER is in figure 4. Altogether the diagram of LAC is in figure 5.

```
Model
Model: "model_1"
_________________________________________________________________
Layer (type)                 Output Shape              Param #
=================================================================
input_1 (InputLayer)         [(None, 6502, 22)]        0
lstm_1612 (LSTM)             (None, 6502, 128)         77312
dense_3224 (Dense)           (None, 6502, 64)          8256
dense_3225 (Dense)           (None, 6502, 22)          1430
dense_3226 (Dense)           (None, 6502, 64)          1472
lstm_1613 (LSTM)             (None, 6502, 128)         98816
Total params: 187,286
Trainable params: 187,286
Non-trainable params: 0
```

**Figure 4. Description of LSTM AUTOENCODER.**

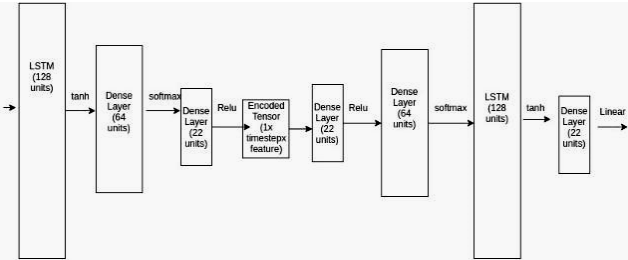

**Figure 5. Proposed LSTM AUTOENCODER model for anomaly detection.**

## 4 EXPERIMENT

### 4.1 System Specification

We used UBUNTU 19.04 64-bit Kernel Linux 5.0.0-37-generic x86_64 MATE 1.20.4 OS which has 502.6 GB of memory with Intel Xeon(R) Gold 6138 CPU @ 2.00Ghz x 80, 4 GPU components(GeForce RTX 2080i) with CUDA version 10.1, driver version 430.50 with 6 TB external disk space. We used R-studio with R-version 3.6.1, Keras(version 2.2.5.0) and Tensorflow(version 2.0.0).

### 4.2 Data Description

We use CERT v6.2 dataset to test LAC. It is an open source database provided by Carnegie Mellon University. The size of the dataset is 101.4 GB. It consists of 18 months worth of network activities of employees in a hypothetical situation. The description of [6] is used to build the dataset. The description of the data is given in the next subsection in three parts -

**4.2.1 Employee Description.** There are 4000 employees in total. They are divided in 46 roles. They all have their own organization email-id irrespective of the role. Some of them changed the organization at some point of the 18 months. They are 362 in total. We did not discard these employees because they might be potential anomalous employees.

The month to month attendance description of the employees is in table 1.

**4.2.2 Raw Log-Line Description.** The main part of the dataset the network activity log-lines are divided into five files, The description of the files are given in table 2. There are decoy.csv and psychometric.csv files too. We used de-coy.csv file to detect the malicious file-tree in the "file.csv" file. The log-lines has There are 21 different activities that we extracted from the log-lines. They are described in table 3. "logon.csv" file has two activities - logon and logoff, "email.csv" file has two activities send and view, "file.csv" has twelve activities they ac-cording to the form "FILE COPY/DELETE/OPEN/WRITE_from removable media_to removable media_decoyed or not" ( every field divided by "_" is binary), "http.csv" has three activities www download, upload, visit.

**Table 1. Monthly attendance of the employees**

| Month   | Attendance |
|---------|------------|
| 2009-12 | 4000       |
| 2010-01 | 4000       |
| 2010-02 | 3977       |
| 2010-03 | 3952       |
| 2010-04 | 3930       |
| 2010-05 | 3911       |
| 2010-06 | 3894       |
| 2010-07 | 3870       |
| 2010-08 | 3853       |
| 2010-09 | 3853       |
| 2010-10 | 3833       |
| 2010-11 | 3803       |
| 2010-12 | 3773       |
| 2011-01 | 3719       |
| 2011-02 | 3697       |
| 2011-03 | 3674       |
| 2011-04 | 3654       |
| 2011-05 | 3639       |

**Table 2. Description of the log-line files (.csv)**

| File-name (#log-lines) | Name of the features |
|------------------------|----------------------|
| Email (10994957)       | Id, date, employee, PC, to, cc, bcc, from, activity, size, attachments, contents |
| File (2014883)         | Id, date, employee, PC, filename, activity, to, removable_media, from_removable_media, content |
| http(117025216)        | Id, date, employee, PC, url, activity, content |
| Device(1551828)        | Id, date, employee, PC, file-tree, activity |
| Logon(3530285)         | Id, date, employee, PC, activity |
| Total(135,117,169)     |                      |

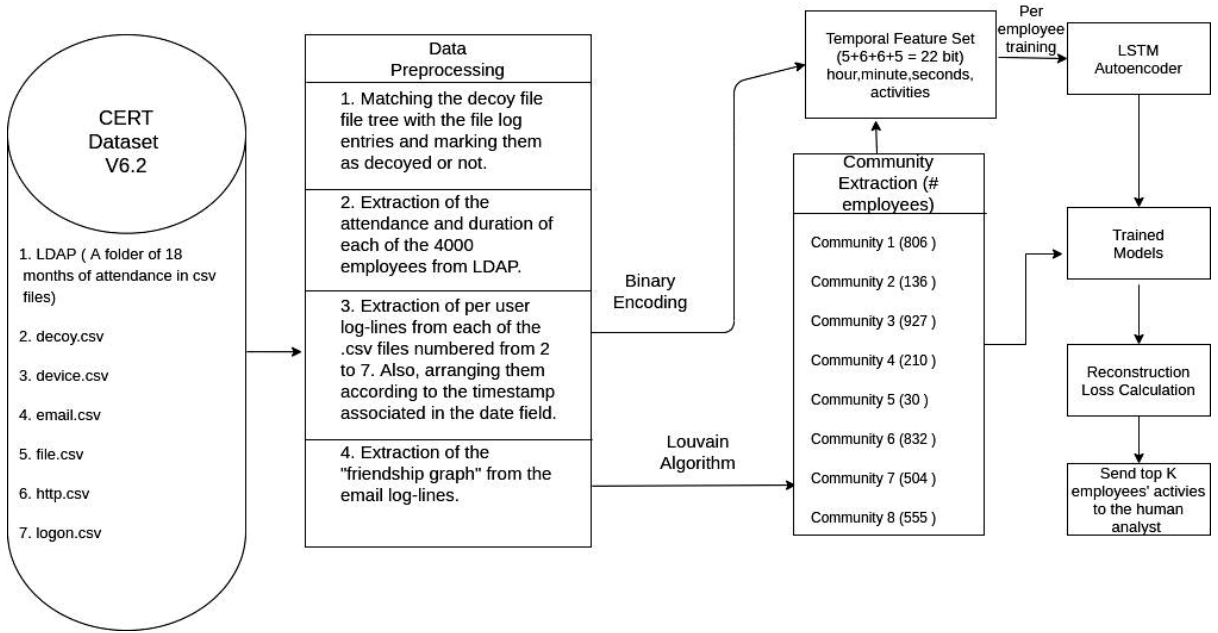

**Figure 6. Model description of LSTM AUTOENCODER using Community detection (LAC)**

**4.2.3 Community Description.** Using Louvain Algorithm we detected eight communities. There are no overlapping of employees among these communities. The description of the communities is given in the Table 5.

## 4.3 Experiment Methods and Description

The aim of our experiments is -
- To detect the potential anomalous employees from the 4000 employee data.
- To test whether the anomalous employees are easier to detect in a community or in absence of it.

We took the text analysis or natural language processing approach for the experiment.

**4.3.1 Data Pre-processing.** In this two phased experi-ment a huge part of time went into data pre-processing. We used R programming language to code the whole pre-processing, Louvain algorithm and LSTM AUTOENCODER and post- processing. The packages we used are in table 4 -

We extracted per employees' log-lines from five files (email.csv, file.csv, http.csv, logon.csv, device.csv) separately, then merged and sorted them according to the date-timestamp and com-munity description in table ??. The main question at this point was to what should be our choice of features? [16] has already shown that the categorical values in the feature does no difference in the anomaly detection. That's why we don't add the categorical values provided in the dataset like role, project, functional unit, department, team, supervisor etc. to our choice of features. Also, according to the metadata provided with the dataset the date and activity features are present among all the files and have the most weight. So, we made our feature in the form of - hours, minutes, seconds, activities for a day. After that we binarised the respective integer value for the four features. The size of one vector becomes 22 bits in this way (5 + 6 + 6 + 5 = 22 bits).

**Table 3. Description of the activities**

| Activity | Activity number |
|---|---|
| Logoff | 1 |
| Logon | 2 |
| Connect | 3 |
| Disconnect | 4 |
| Send | 5 |
| View | 6 |
| File Copy_0_1_0 | 7 |
| File Copy_0_1_1 | 8 |
| File Copy_1_0_0 | 9 |
| File Copy_1_0_1 | 10 |
| File Delete_0_1_0 | 11 |
| File Delete_0_1_1 | 12 |
| File Open_0_0_0 | 13 |
| File Open_0_0_1 | 14 |
| File Open_0_1_0 | 15 |
| File Open_0_1_1 | 16 |
| File Write_1_0_0 | 17 |
| File Write_1_0_1 | 18 |
| WWW Download | 19 |
| WWW Upload | 20 |
| WWW Visit | 21 |

**Table 4. Description of the packages in R for data pre-processing and writing the model**

| Package name | Usage |
|---|---|
| Keras | To code the LSTM AUTOENCODER |
| data.table | To read the .csv files |
| Tidyverse | To separate, unite, and further dividing the strings. |
| Igraph | To extract the communities. |
| Sqldf | Sql query to extract data |
| R.utils | Integer to binary conversion |
| Reticulate | To run keras on top of python surface in R |
| muStat | To calculate min, max, mean and standard deviation |
| Lubridate | To convert character to Date object. |
| Readr | To read and copy a bulk number of files at once |
| Ggplot2 | To visualize the result |

**4.3.2 Experiment.** After the data pre-processing part we trained and stored the LSTM AUTOENCODER individually for every employees with 70% of the binarized data that had been extracted and sorted accordingly. The hyper parameters we used in training are - loss function : mean squared error, optimizer : adam, learning rate : 0.01, decay : 0.01 and metrics : "accuracy". We then tested the stored models with the remaining 30% data, where we took account of the re-construction loss. There were 4000 models in total. Then we tested those models individually again against the com-munity data which is basically the binarised, interleaved and sequenced log-lines of all the employees altogether in a community. We then retrieved the log-lines again for that employee from this whole lot of different log-lines, with higher reconstruction loss for anomaly employees and lower reconstruction loss for normal employees. The community, analysis and its evaluation is discussed in the next section 4.4. The description for the whole process is given in figure 6.

## 4.4 Result, Analysis and Evaluation

**4.4.1 Result.** In the CERT dataset V6.2 there were FIVE situations as answer that indicated FIVE employees as anomalous. We used them as ground truth to test our results. We successfully detected all the anomalous employees using our approach with many false positives included. The results are given in figure 7

**4.4.2 Analysis.** We did our analysis in two phases. In the first phase we tested our hypothesis on the community approach. LACs per employee were already trained by the "normal" behavior of those employees. So, when we were again training them with the interleaved and sequentially arranged data for a whole community the anomalous employees were expected to show a larger reconstruction loss. LAC showed it successfully. If we pass all the top-K (budget of K will be decided by the organization where k is the number of employees to be sent to the human analyst, suppose 5) Results to a human analyst from each community then there will be assurance that the potential anomalous employees would obviously be there. There was one anomalous employee each in community 1, 6 and 7 and 2 anomalous employees in community 3. We also calculate the standard deviation of the normalized loss per community which is in table 5.

**Table 5. Description of the Standard deviation and number of employees in every community**

| Community # | Employee# | Standard Deviation |
|---|---|---|
| 1 | 806 | 2.19577041239726E-05 |
| 2 | 136 | 7.83589986017541E-05 |
| 3 | 927 | 1.23405734890555E-05 |
| 4 | 210 | 5.45469004617978E-05 |
| 5 | 30 | 0.00036854479559 |
| 6 | 832 | 1.36956941822872E-05 |
| 7 | 504 | 2.27847603486002E-05 |
| 8 | 555 | 1.47179483456863E-05 |

The data from the table shows that the more the anomalous employees in a community reconstruction loss are supposed to be closer which actually proves our community approach.

In the second phase we randomly picked 999 employees from the 4000 employees and pushed the anomalous employee from community 6 and did the same experiment again on those 1000 employees. The result is in figure 8. The experiment result supports our hypothesis again. All the code for data pre-processing, model training and analysis will be found in this link - https://github.com/smlab-niser/LAC

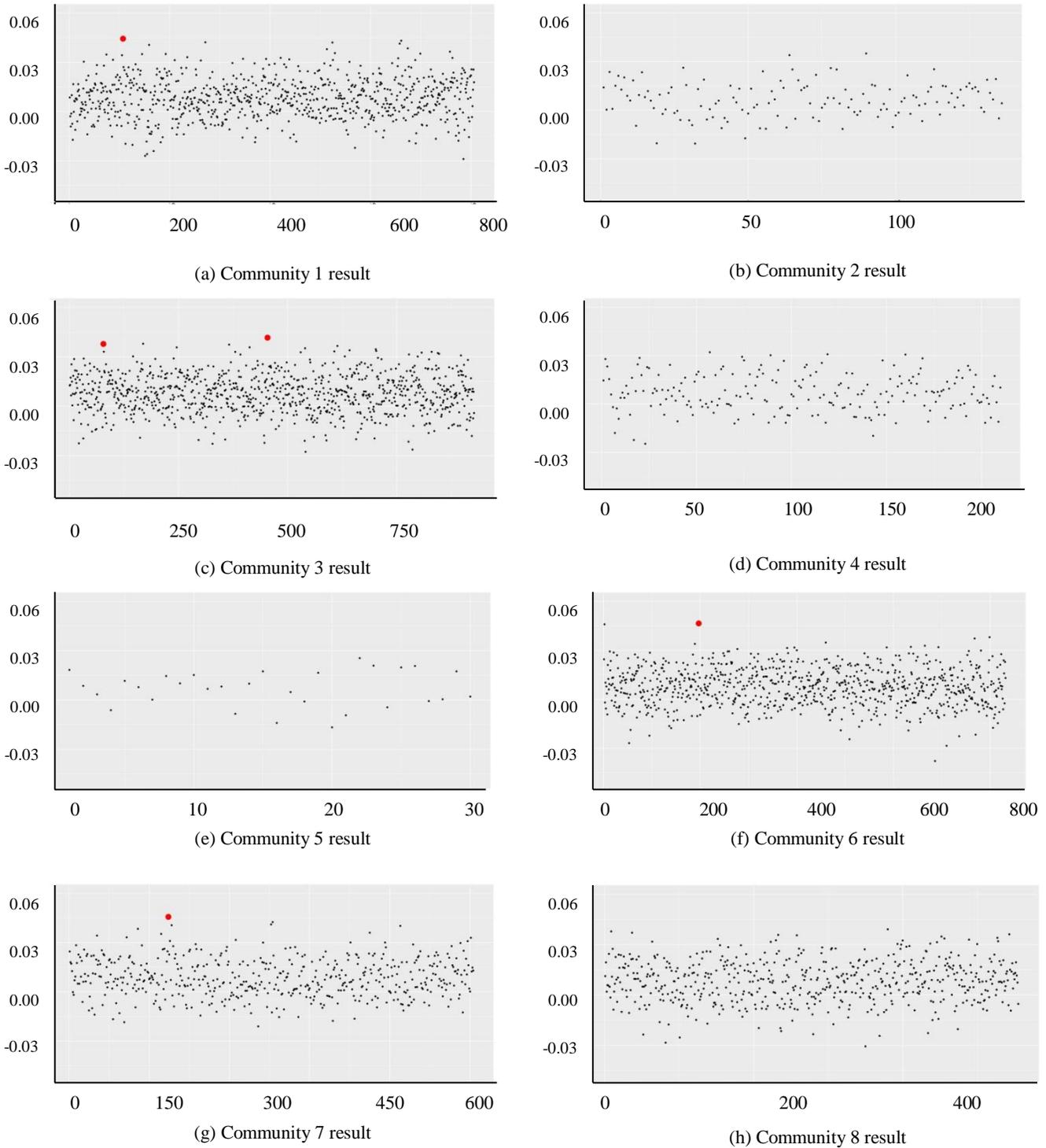

**Figure 7. Images of Anomalies in Communities in comparison to Ground Truth where, Anomalous employees are highlighted by red color. The x-axis represents the loss and y-axis represents the ID indexes of the users.**

## 5 CONCLUSION

We proposed LAC, a deep learning model using LSTM AUTOENCODER that is capable of reproducing the action sequences again with less reconstruction loss for the normal employees and greater reconstruction loss for the anomalous ones, with the help of its inherent capability to store temporal behavior in LSTM cell. We have tested LAC against one of the most widely used and standard dataset CERT V6.2.

Our novel approaches here are - Usage of community of employees instead of analysis on a whole set, Usage of LSTM AUTOENCODER to model the deep learning approach, Build-ing feature set in the granularity of event occurrence with respect to activities which gives us the ability to detect the anomalous employee more accurately. Our performance with respect to time is also quite encouraging (data pre-processing time around 9 minutes in average per employee and training time about 1 minute in average per employee). On the basis of this, we believe that LAC has a great potential to perform even better in lightweight devices. In future we want to extend LAC to be able to exe-cute different community detector algorithms, in different feature set (e.g. email content, web visit activities, file tree hierarchy etc.) to detect the underlying "friendship graph" and hence to build the communities efficiently where the features might be detected on its own too.

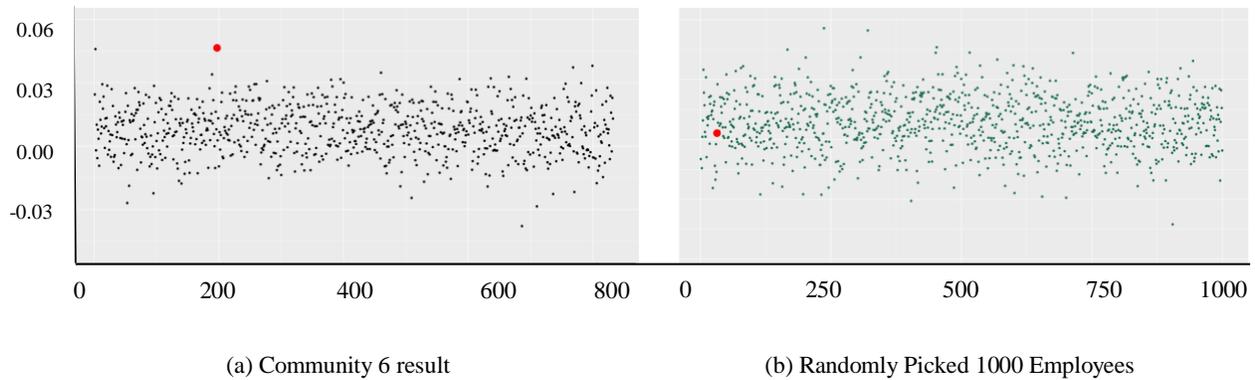

**Figure 8. Comparison of Anomalies between randomly picked 1000 employees with Community 6 where, Anomalous employees are highlighted by red color. The x-axis represents the loss and y-axis represents the ID indexes of the users.**